\documentclass[twocolumn,pra,superscriptaddress,longbibliography]{revtex4-1}
\usepackage{graphicx,amsmath,amssymb}
\usepackage[colorlinks=true, citecolor=blue, urlcolor=blue, linkcolor=red]{hyperref}
\begin{document}
\title{Continuous-variable quantum sensing of a dissipative reservoir}
\author{Yi-Da Sha}
\affiliation{Key Laboratory of Theoretical Physics of Gansu Province, \\
and Lanzhou Center for Theoretical Physics, Lanzhou University, Lanzhou, 730000, China}
\author{Wei Wu}
\email{wuw@lzu.edu.cn}
\affiliation{Key Laboratory of Theoretical Physics of Gansu Province, \\
and Lanzhou Center for Theoretical Physics, Lanzhou University, Lanzhou, 730000, China}

\begin{abstract}
We propose a continuous-variable quantum sensing scheme, in which a harmonic oscillator is employed as the probe to estimate the parameters in the spectral density of a quantum reservoir, within a non-Markovian dynamical framework. It is revealed that the sensing sensitivity can be effectively boosted by (i) optimizing the weight of the momentum-position-type coupling in the whole probe-reservoir interaction Hamiltonian, (ii) the initial quantum squeezing resource provided by the probe, (iii) the noncanonical equilibration induced by the non-Markovian effect, and (iv) applying an external driving field. Our results may have some potential applications in understanding and controlling the decoherence of dissipative continuous-variable systems.
\end{abstract}
\maketitle

\section{Introduction}\label{sec:sec1}

Quantum sensing aims at characterizing, measuring and estimating an unknown parameter of interest with ultra high sensitivity, which can surpass the standard bound set by classical statistics, with the help of the so-called quantum superiority~\cite{RevModPhys.89.035002,RevModPhys.90.035005,doi:10.1116/1.5119961}. Such a quantum superiority is commonly established by employing certain quantum resources, such as quantum entanglement~\cite{Lachance-Quirion425,PhysRevLett.122.123605,Unternahrer:18,Zou6381,Nagata726,PhysRevLett.124.060402,yamamoto2022errormitigated}, quantum squeezing~\cite{PhysRevD.23.1693,PhysRevLett.127.160501,PhysRevLett.119.193601,PhysRevLett.123.113602,PhysRevLett.123.231107,PhysRevLett.121.110505} as well as quantum criticality~\cite{PhysRevA.80.012318,PhysRevA.78.042106,PhysRevA.78.042105,PhysRevLett.124.120504,PhysRevLett.126.010502,PhysRevLett.121.020402}, which have no counterparts in classical physics. Quantum sensing has been widely applied to the studies of various quantum thermometers~\cite{PhysRevX.10.011018,PhysRevResearch.3.043039,PhysRevResearch.2.033394,PhysRevA.96.062103,PhysRevA.97.063619,PhysRevLett.114.220405} and quantum magnetometries~\cite{PhysRevLett.126.070503,PhysRevLett.127.193601,PhysRevLett.122.173601,PhysRevA.99.062330}.

Recently, much attention has been focused on the sensing of a quantum reservoir, in which a probe is used to indirectly measure the details about the spectral density of the quantum reservoir~\cite{PhysRevApplied.15.054042,PhysRevA.97.012126,PhysRevA.97.012125,SALARISEHDARAN2019126006,PhysRevA.101.032112,Tamascelli_2020,PhysRevA.102.012223,PhysRevLett.98.160401,PhysRevA.92.010302}. It is generally believed that the spectral density fully characterizes the frequency dependence of the interaction strengths as well as the dispersion relation of a quantum reservoir. In the theory of open quantum systems, the spectral density plays a crucial role in determining the decay rates of dissipation and decoherence~\cite{Breuer,Weiss,RevModPhys.59.1,RevModPhys.89.015001,RevModPhys.88.021002}. Moreover, in many dynamical control schemes of decoherence reduction, say, the strategy of decoupling pulses in Refs.~\cite{PhysRevLett.87.270405,PhysRevLett.98.100504,PhysRevLett.101.010403}, prior knowledge about the spectral density is indispensable. Unfortunately, the spectral density itself is not a physical observable and is usually described by a set of phenomenological parameters, which cannot be derived from first principles. Thus, quantum sensing of the spectral density is of great scientific significance from the perspectives of understanding and controlling the decoherence.

Several sensing schemes of the spectral density have been proposed in previous works~\cite{PhysRevApplied.15.054042,PhysRevA.97.012126,PhysRevA.97.012125,SALARISEHDARAN2019126006,PhysRevA.101.032112,Tamascelli_2020,PhysRevA.102.012223,PhysRevLett.98.160401,PhysRevA.92.010302,jonsson2022gaussian}. However, almost all these existing studies have restricted their attentions to (i) the discrete-variable probe case~\cite{PhysRevApplied.15.054042,PhysRevA.97.012126,SALARISEHDARAN2019126006,PhysRevA.101.032112,Tamascelli_2020,PhysRevA.92.010302}, i.e., the probe is made of a qubit or a few-level system, or (ii) the continuous-variable probe case within a Markovian approximate dynamical framework~\cite{PhysRevA.97.012125,PhysRevA.102.012223,PhysRevLett.98.160401,PhysRevA.95.012109}. Compared with the qubit-based implementations, the continuous-variable settings have some special features: the so-called unconditionalness, which improves their efficiencies in certain quantum tasks~\cite{RevModPhys.77.513}. On the other hand, it has been demonstrated that non-Markovianity can effectively boost the precision of a qubit-based parameter estimation protocol~\cite{PhysRevApplied.15.054042,PhysRevA.102.032607,PhysRevLett.109.233601,PhysRevLett.127.060501,PhysRevA.88.035806}. Very few studies focus on the non-Markovian effect on the performance of a continuous-variable-based sensing scenario. In this sense, going beyond the usual limitation of a  Markovian approximation is highly desirable for the continuous-variable sensing of a quantum reservoir.

To address the above concerns, in this paper, we propose a continuous-variable quantum sensing scheme, employing a harmonic oscillator as the probe, to estimate the parameters of the spectral density of a quantum reservoir. The influences of the initial quantum squeezing, the probe-reservoir coupling type and the noncanonical equilibrium state induced by the non-Markovianity on the sensing performance are investigated. Moreover, we reveal that an external driving field, which is solely applied to the probe, can be used as a dynamical tool to improve the sensing performance.

This paper is organized as follows. In Sec.~\ref{sec:sec2}, we briefly outline a basic formalism about the quantum sensing and propose our scheme in details. In Sec.~\ref{sec:sec3}, we report our main results. The conclusion of this paper is drawn in Sec.~\ref{sec:sec4}. In the three appendices, we provide some additional materials about the main text. Throughout the paper, for the sake of simplicity, we set $\hbar=k_{\mathrm{B}}=m=1$, where $m$ denotes the mass of the harmonic oscillator probe in our scheme, and the inverse temperature is accordingly re-scaled as $\beta=1/T$.

\section{Non-Markovian sensing}\label{sec:sec2}

\subsection{Quantum Fisher information (QFI)}\label{sec:sec2a}

In a conventional quantum sensing protocol, one needs a quantum probe, which is initially prepared in a certain state $\rho_{\mathrm{in}}$, and one couples it to the target system, which contains the parameter of interest $\lambda$. Due to the interaction between the probe and the target system, the information about $\lambda$ is then encoded into the state of the probe via a mapping $\rho_{\lambda}=\mathcal{M}_{\lambda}(\rho_{\mathrm{in}})$. Here, the $\lambda$-dependent superoperator $\mathcal{M}_{\lambda}$ can be physically realized by either a unitary~\cite{Hauke2016,PhysRevB.99.045117,Oh2019} or a nonunitary encoding process~\cite{PhysRevApplied.15.054042,PhysRevA.97.012126,PhysRevA.97.012125,SALARISEHDARAN2019126006,PhysRevA.101.032112,Tamascelli_2020,PhysRevA.102.012223,PhysRevLett.98.160401,PhysRevA.92.010302}. Next, by measuring a certain physical observable with respect to $\rho_{\lambda}$, an estimator $\hat{\lambda}$ can be constructed. The uncertainty of $\hat{\lambda}$ is constrained by the famous quantum Cram\'{e}r-Rao bound~\cite{Liu_2019}
\begin{equation}\label{eq:eq1}
\delta^{2}\hat{\lambda}\geq \frac{1}{\upsilon\mathcal{F}_{\lambda}},
\end{equation}
where $\delta\hat{\lambda}$ is the standard error of the estimator, $\upsilon$ denotes the repeated measurement times, and $\mathcal{F}_{\lambda}\equiv\text{Tr}(\rho_{\lambda}\varsigma^2)$ with $\varsigma$ defined by $\partial_\lambda\rho_{\lambda}=\frac{1}{2}(\varsigma\rho_\lambda+\rho_\lambda \varsigma)$ is the quantum Fisher information (QFI). The QFI characterizes the statistical information about $\lambda$ included in $\rho_{\lambda}$ and is independent of the selected measurement scenario. From Eq.~(\ref{eq:eq1}), one can conclude that a larger QFI corresponds to a higher sensing precision.

In the study of quantum parameter estimation, researchers are interested in the scaling relation, which describes the connection between the QFI $\mathcal{F}_{\lambda}$ and the number of quantum resources $\bar{n}$ contained in $\rho_\text{in}$. The scaling relation is one of the most important indexes to evaluate the performance of a quantum sensing scheme. If $\mathcal{F}_{\lambda}$ is proportional to $\bar{n}$, the scaling relation is called the standard quantum limit (SQL). It has been revealed that the SQL can be surpassed by using certain quantum resources. For example, by employing the quantum squeezing, the scaling relation in a Mach-Zehnder interferometer can be boosted to the Zeno limit in the absence of noise~\cite{PhysRevD.23.1693,PhysRevLett.123.040402}. In this paper, going beyond the above noiseless situation, we shall investigate the influences of quantum squeezing on the sensing performance in the presence of decoherence, which is inevitably generated by the probe-reservoir interaction.

\subsection{The QFI of a Gaussian state}\label{sec:sec2b}

If the probe is a Gaussian continuous-variable system, its quantum state $\rho_{\lambda}$ can be fully characterized by the first two moments $\pmb{d}$ (the displacement vector) and $\pmb{\sigma}$ (the covariant
matrix)~\cite{RevModPhys.77.513}. Defining the quadrature operator as $\pmb{Q}=(x,p)^{\text{T}}$, the elements of ${\pmb{d}}$ and ${\pmb \sigma}$ are, respectively, defined by $d_{i} =\text{Tr}(\rho_\lambda Q_{i})$ and $\sigma _{ij} =\text{Tr}[\rho_\lambda \{\Delta Q_{i},\Delta Q_{j}\}]$
with $\Delta Q_{i}=Q_{i}-d_{i}$. With expressions of $\pmb{d}$ and $\pmb{\sigma}$ at hand, the QFI with respect to the Gaussian state $\rho_{\lambda}$ can be calculated as~\cite{_afr_nek_2015,_afr_nek_2018,Gao2014}
\begin{equation}\label{eq:eq2}
\mathcal{F}_{\lambda}=\frac{1}{2}[\text{vec}(\partial_\lambda {\pmb \sigma})]^\dag\pmb{M}^{-1}\text{vec}(\partial_\lambda {\pmb \sigma})+2(\partial_{\lambda}\pmb{d})^{\dagger}\pmb{\sigma}^{-1}\partial_{\lambda}\pmb{d},
\end{equation}
where $\text{vec}(\cdot)$ denotes the vectorization of a given matrix, and $\pmb{M}={\pmb\sigma}\otimes{\pmb\sigma}-\pmb{\varpi}\otimes \pmb{\varpi}$ with $[Q_{i},Q_{j}]=i\varpi_{ij}$.

\subsection{Our sensing scheme}\label{sec:sec2c}

\begin{figure}
\includegraphics[angle=0,width=0.45\textwidth]{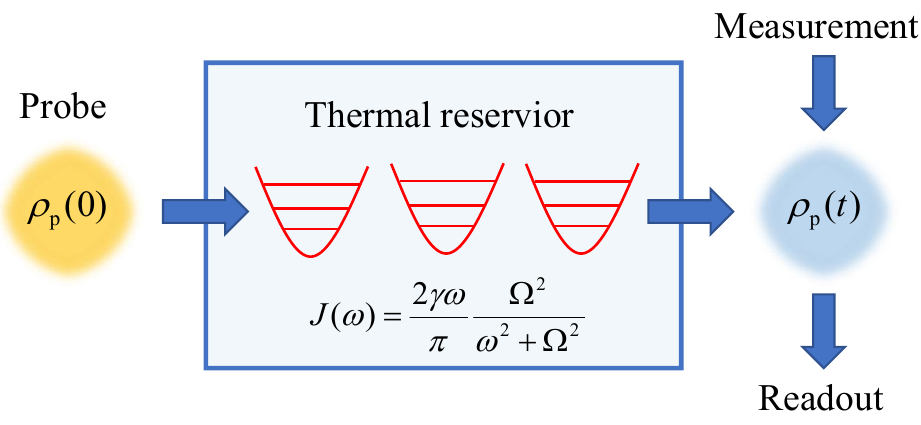}
\caption{\label{fig:fig1} Schematic diagram of our quantum sensing scheme. A continuous-variable probe, which is initially prepared in a squeezed state, is employed to reveal the details of the spectral density $J(\omega)$ of a thermal reservoir, which consists of an infinite collection of noninteracting harmonic oscillators.}
\end{figure}

As displayed in Fig.~\ref{fig:fig1}, in our quantum sensing scheme, a harmonic oscillator is employed as the probe to detect the spectral density of a dissipative thermal reservoir. The whole Hamiltonian of the probe plus the reservoir can be described by $H=H_{\text{p}}+H_{\text{r}}+H_{\text{int}}+H_{\text{c}}$. Here
\begin{equation}
H_{\text{p}}=\frac{p^{2}}{2m}+\frac{1}{2}m\omega_{0}^{2}x^{2}
\end{equation}
is the Hamiltonian of the probe,
\begin{equation}
H_{\text{r}}=\sum_{k}\bigg{(}\frac{p^{2}_{k}}{2m_{k}}+\frac{1}{2}m_{k}\omega_{k}^{2}x_{k}^{2}\bigg{)}
\end{equation}
denotes the Hamiltonian of the thermal reservoir,
\begin{equation}
H_{\text{int}}=\sum_{k}c_{k}x_{k}\mathcal{S}
\end{equation}
describes the probe-reservoir interaction, which is responsible for the encoding process in our sensing scheme, and
\begin{equation}
H_{\text{c}}=\sum_{k}\frac{c_{k}^{2}}{2m_{k}\omega_{k}^{2}}\mathcal{S}^{2}
\end{equation}
is the so-called counter-term~\cite{Breuer}, which compensates for the frequency shift induced by the interaction between the probe and the thermal reservoir~\cite{Breuer,PhysRevD.45.2843,PhysRevD.53.2012,PhysRevA.95.052109}. The quantities $\{x,p,m,\omega_{0}\}$ and $\{x_{k},p_{k},m_{k},\omega_{k}\}$ are the position, the momentum, the mass, and the frequency of the probe and the $k$th harmonic oscillator of the reservoir, respectively. Parameters $c_{k}$ quantify the probe-reservoir coupling strengths, and $\mathcal{S}$ denotes the coupling operator.

In this paper, we consider the following general coupling operator
\begin{equation}
\mathcal{S}=x\cos\theta+p\sin\theta.
\end{equation}
By varying the coupling angle $\theta$, both position-position-type ($xx_{k}$-type) coupling and momentum-position-type ($px_{k}$-type) coupling can be taken into account. Notably, when $\theta=0$, the standard Caldeira-Leggett model~\cite{Breuer,PhysRevD.45.2843,PhysRevD.53.2012,PhysRevA.95.052109} can be recovered. In Refs.~\cite{PhysRevA.96.012109,PhysRevA.95.052109}, the authors reported that the momentum-position-type coupling can remarkably modify the dissipation experienced by the probe. Their results inspire us to explore the effect of the momentum-position-type coupling on the sensitivity of the quantum sensing.

The spectral density in our model, which is defined by
\begin{equation}
J(\omega)\equiv\sum_{k}\frac{c_{k}^{2}}{2m_{k}\omega_{k}}\delta(\omega-\omega_{k}),
\end{equation}
fully determines the properties of the thermal reservoir.
In the following, we assume $J(\omega)$ is an Ohmic spectral density with a Lorentz-Drude cutoff
\begin{equation}
J(\omega)=\frac{2\gamma\omega}{\pi}\frac{\Omega^{2}}{\omega^{2}+\Omega^{2}},
\end{equation}
where $\gamma$ (the coupling strength) and $\Omega$ (the cutoff frequency) are the two parameters to be estimated in this paper, namely $\lambda=\gamma$ or $\Omega$ henceforth. The Ohmic-type spectral density constitutes a very general form to describe many different types of reservoirs. By varying the scopes of the coupling strength $\gamma$ and the cutoff frequency $\Omega$, the Ohmic-type spectral density can be employed to simulate the dynamics of charged interstitials in metals~\cite{RevModPhys.59.1,RevModPhys.89.015001}. Moreover, an Ohmic model with a Lorenz-Drude regularization can be used to describe the electronic energy transfer dynamics in photosynthetic a pigment-protein complex and the Fenna-Matthews-Olson complex~\cite{doi:10.1073/pnas.0908989106}. On the other hand, the choice of the Lorentz-Drude-type Ohmic spectral density can greatly reduce the difficulties in deriving the expression of the output state. Thus, due to the above two reasons, we choose the Ohmic spectral density as the example to demonstrate the feasibility of our proposed sensing scheme.

The probe is initially prepared in a squeezed state
\begin{equation}\label{eq:eq10}
\rho_{\text{p}}(0)=S(r)D(\alpha)|0\rangle\langle 0|D^\dag(\alpha)S^\dag(r),
\end{equation}
where $D(\alpha)\equiv\exp(\alpha a^{\dagger}-\alpha^{*}a)$ and $S(r)\equiv\exp[r(a^{2}-a^{\dag2})/2]$ are the displacement and the squeeze operators, respectively. Here, $a|0\rangle=0$ denotes the Fock vacuum state and $a\equiv(x+ip)/\sqrt{2}$ is introduced as the annihilation operator of the probe. It is easy to prove that $\rho_{\text{p}}(0)$ is a Gaussian state. Assuming the whole initial state of the probe plus the reservoir is $\rho_{\text{p}}(0)\otimes\rho_{\text{r}}^{\text{G}}$ with $\rho_{\text{r}}^{\text{G}}=e^{-\beta H_{\text{r}}}/\text{Tr}(e^{-\beta H_{\text{r}}})$ being the canonical Gibbs state of the thermal reservoir, the dissipative dynamics generated by $H$ can be exactly derived by using the quantum master equation approach. As demonstrated in Refs.~\cite{Breuer,PhysRevA.96.012109,PhysRevA.95.052109,PhysRevA.102.022228}, the Gaussianity of the probe can be fully preserved during the time evolution as a consequence of the bilinear structure of the global Hamiltonian $H$. On the other hand, using the Heisenberg equation of motion, the expressions of $\pmb{d}$ and $\pmb{\sigma}$ can be obtained (see Appendices A and B for details). As long as the first two momenta are obtained, the QFI can be accordingly computed by making use of Eq.~(\ref{eq:eq2}).

When performing the numerical calculations to $\mathcal{F}_{\lambda}$ using Eq.~(\ref{eq:eq2}), one needs to handle first-order derivatives,  $\partial_{\lambda}\pmb{d}$ and $\partial_{\lambda}\pmb{\sigma}$ in Eq.~(\ref{eq:eq2}). In this paper, the first-order derivative for an arbitrary $\lambda$-dependent function $f_{\lambda}$ is numerically done by adopting the following finite difference method (see Appendix C and Ref. \cite{PhysRevB.89.134101} for more details):
\begin{equation}\label{eq:eq8}
\partial_{\lambda} f_{\lambda}\simeq\frac{-f_{\lambda+2\delta}+8f_{\lambda+\delta}-8f_{\lambda-\delta}+f_{\lambda-2\delta}}{12\delta}.
\end{equation}
We set $\delta/\lambda=10^{-6}$, which provides a very good accuracy.

\section{Results}\label{sec:sec3}

\subsection{Non-Markovian dynamics of the QFI}\label{sec:sec3a}

\begin{figure}
\includegraphics[angle=0,width=0.425\textwidth]{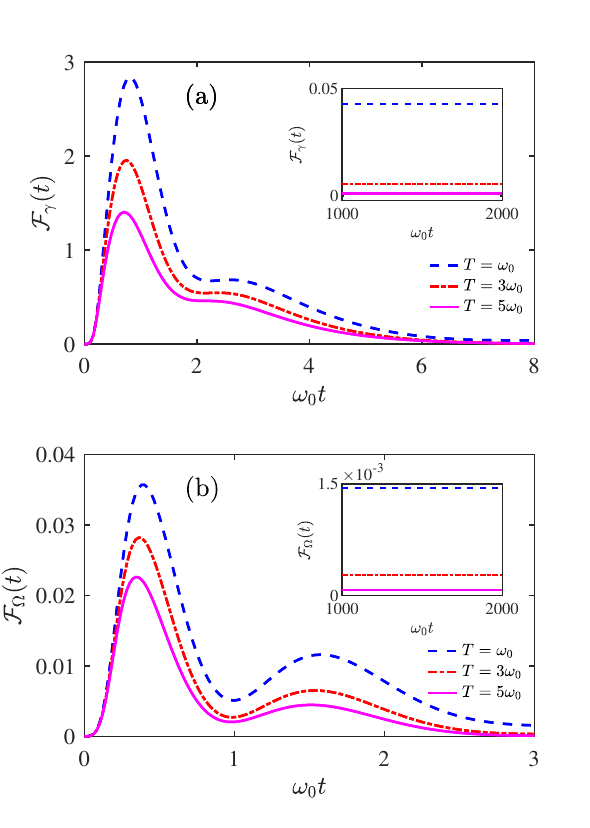}
\caption{\label{fig:fig2} The non-Markovian dynamics of (a) $\mathcal{F}_{\gamma}(t)$ and (b) $\mathcal{F}_{\Omega}(t)$ with different temperatures: $T=\omega_{0}$ (blue dashed lines), $T=3\omega_{0}$ (red-dotdashed lines), and $T=5\omega_{0}$ (magenta solid lines). The insets depict the steady-state QFI $\mathcal{F}_{\lambda}(\infty)$ in the long-time regime. One can see that the value of $\mathcal{F}_{\lambda}(\infty)$ is small, but still positive. Other parameters are chosen as $\omega_0=0.5$ THz, $\Omega=10\omega_{0}$, $\gamma=3\omega_{0}$, $r=5\omega_0$, $\alpha=0$ and, $\theta=0$.}
\end{figure}

\begin{figure}
\includegraphics[angle=0,width=0.485\textwidth]{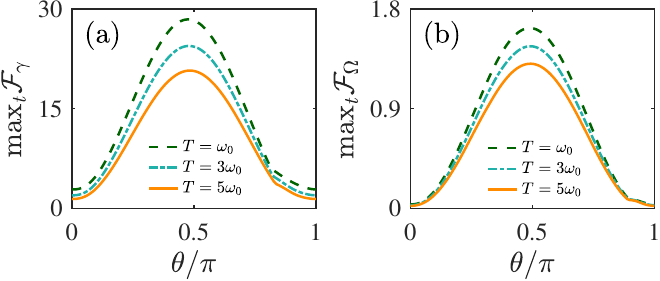}
\caption{\label{fig:fig3} The maximal QFI of (a) $\max_{t}\mathcal{F}_{\gamma}$ and (b) $\max_{t}\mathcal{F}_{\Omega}$ are plotted as a function of the coupling angle $\theta$ with different temperatures: $T=\omega_{0}$ (green dashed lines), $T=3\omega_{0}$ (cyan-dotdashed lines), and $T=5\omega_{0}$ (orange solid lines). Due to the rotational symmetry of $\mathcal{S}$, we here restrict our study to $\theta\in[0,\pi]$. Other parameters are the same as those of Fig.~\ref{fig:fig2}.}
\end{figure}

The non-Markovian dynamical behavior of the QFI is displayed in Fig.~\ref{fig:fig2}. At the beginning, no message about the spectral density is included in the initial state $\rho_{\text{p}}(0)$, leading to $\mathcal{F}_{\lambda}(0)=0$. As the encoding time becomes longer, the probe-reservoir interaction generates the information of the spectral density, which results in the increase of the QFI. After arriving at a maximum value,  $\mathcal{F}_{\lambda}(t)$ begins to decrease as a result of decoherence. Finally, as the probe evolves to its steady state $\rho_{\text{p}}(0)$ in the long-encoding-time limit, the value of $\mathcal{F}_{\lambda}(t)$ remains unchanged. In the inserts of Fig.~\ref{fig:fig2}, we plot the QFI in the long-encoding-time regime. One can clearly see $\mathcal{F}_{\lambda}(\infty)>0$, which is induced by the purely non-Markovian effect and is evidently different from that of the previous Markovian case~\cite{PhysRevA.97.012125,PhysRevA.102.012223,PhysRevLett.98.160401,PhysRevA.95.012109}. More detailed discussions on the behavior of $\mathcal{F}_{\lambda}(\infty)$ are presented in the Sec.~\ref{sec:sec3c}.

The above result means there exists an optimal encoding time which can maximize the QFI. The occurrence of such a maximal QFI with respect to the optimal encoding time originates from the competition between the indispensable encoding and the unavoidable decoherence~\cite{PhysRevApplied.15.054042,PhysRevResearch.3.043039}, which are induced by the probe-reservoir interaction. In Fig.~\ref{fig:fig3}, we plot the maximal QFI, $\max_{t}\mathcal{F}_{\lambda}$, as a function of the coupling angle $\theta$. One can find that the maximal QFI can be further improved by adjusting the coupling angle. This result means the pure position-position-type coupling in the standard Caldeira-Leggett model is not the prime choice for obtaining the maximum sensing precision. Via adding the momentum-position-type coupling in $H_{\text{sb}}$, one can design the most efficient probe-reservoir interaction Hamiltonian for the encoding process.

Moreover, we observe that the dynamics of QFI can exhibit an oscillating behavior, e.g., the blue dashed line in Fig.~\ref{fig:fig2} (b), which will result in multiple local maxima. This behavior is different from the previous result~\cite{PhysRevA.97.012125} in which there exists one single peak. Such a result was also reported in Refs.~\cite{PhysRevResearch.3.043039,PhysRevA.102.032607} and may be linked to the exchange of information between the thermal reservoir and the probe. The reversed information flow from the reservoir back to the probe is commonly regarded as evidence of non-Markovianity.

\subsection{The scaling relation}\label{sec:sec3b}

\begin{figure}
\includegraphics[angle=0,width=0.485\textwidth]{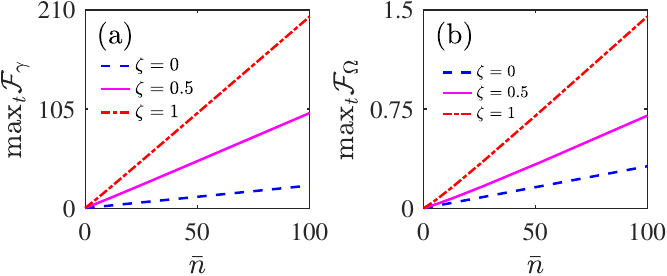}
\caption{\label{fig:fig4} The maximum QFI of (a) $\max_{t}\mathcal{F}_{\gamma}$ and (b) $\max_{t}\mathcal{F}_{\Omega}$ vs the averaged number of photons, $\bar{n}$, with different squeezing ratios: $\zeta=0$ (blue dashed lines), $\zeta=0.5$ (magenta solid lines), and $\zeta=1$ (red dot-dashed lines). Other parameters are chosen as $\omega_0=1$ THz, $\Omega=10\omega_{0}$, $T=3\omega_{0}$, and $\gamma=\omega_{0}$.}
\end{figure}

For the initial state given by Eq.~(\ref{eq:eq10}), the averaged photon number reads
\begin{equation}
\bar{n}\equiv\text{Tr}\big{[}\rho_{\text{p}}(0)a^{\dagger}a\big{]}=|\alpha|^{2}+\sinh^{2}r,
\end{equation}
which can be regarded as the quantum resource employed in our sensing scheme. Furthermore, we introduce a squeezing ratio $\zeta\equiv\sinh^{2}r/\bar{n}$ to quantify the weight of quantum squeezing in the total quantum resource. The ratio of $\zeta$ varies from $\zeta=0$ for a purely coherent state to $\zeta=1$ for a purely squeezed vacuum state. Next, we shall explore the scaling relation versus $\bar{n}$ and $\zeta$.

The scaling relations between $\max_{t}\mathcal{F}_{\lambda}$ and $\bar{n}$ with different squeezing ratios are displayed in Fig.~\ref{fig:fig4}. One can find that $\max_{t}\mathcal{F}_{\lambda}$ is proportional to $\bar{n}$, which means the sensing precision scales as the SQL in our scheme. However, we find  that the slope of the SQL can be enhanced by increasing the weight of quantum squeezing, which implies that quantum squeezing can be used as a resource to boost the sensing performance. These results are in agreement with many previous studies of noisy quantum metrology: in a noiseless ideal case, using quantum resource can indeed boost the metrological performance and result in a better scaling relation (such as the Zeno limit~\cite{PhysRevD.23.1693,PhysRevLett.123.040402} or the Heisenberg limit~\cite{PhysRevLett.79.3865}), but these quantum superiorities generally degrade back to the SQL under the influence of decoherence~\cite{PhysRevApplied.15.054042,Wang_2017,PhysRevLett.120.140501,PhysRevLett.109.233601}. Such a result is called the no-go theorem of noisy quantum metrology.

\subsection{Breakdown of the Markovian approximation in the long-encoding-time regime}\label{sec:sec3c}

\begin{figure}
\includegraphics[angle=0,width=0.485\textwidth]{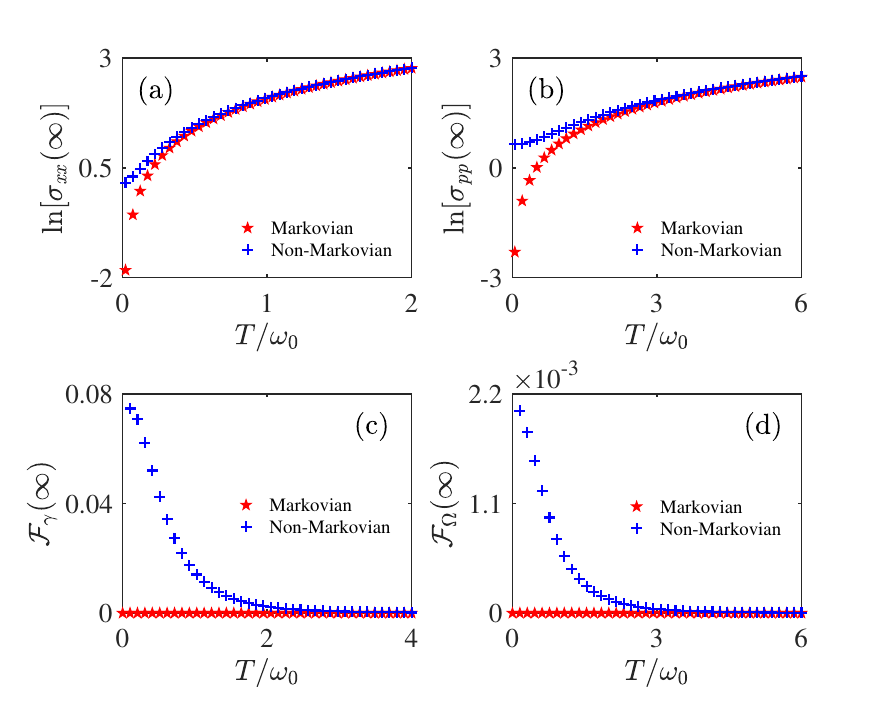}
\caption{\label{fig:fig5} The steady-state second moments of (a) $\sigma_{xx}(\infty)$ and (b) $\sigma_{pp}(\infty)$ are plotted as a function of temperature. The steady-state QFIs of (c) $\mathcal{F}_{\gamma}(\infty)$ and (d) $\mathcal{F}_{\Omega}(\infty)$ are displayed with the increase of temperature. The red five-pointed stars are Markovian results, while the blue crosses represent the results predicted by our non-Markovian method. Other parameters are chosen as $\omega_{0}=0.5~\text{THz}$, $\Omega=10\omega_0$, $\gamma=3\omega_0$, $r=\omega_0$, $\alpha=0$, and $\theta=0$.}
\end{figure}

In this subsection, we shall discuss the steady-state QFI in the long-encoding-time regime. As demonstrated in many previous works~\cite{Breuer,PhysRevE.90.022122,PhysRevA.89.012128,PhysRevA.90.032114,doi:10.1063/1.4722336,Xiong2015ER}, within the Markovian treatment, the long-time steady state of the probe can be described by a canonical Gibbs state at the same temperature as the quantum reservoir. On the other hand, the contribution from the counter term $H_{\text{c}}$ is completely washed out by the probe-reservoir interaction under the Markovian approximation~\cite{Breuer}. The above two points mean the steady state of the probe will be
\begin{equation}\label{eq:eq12}
\rho_{\text{p}}(\infty)=\rho_{\text{p}}^{\text{G}}=\frac{e^{-\beta H_{\text{p}}}}{\text{Tr}(e^{-\beta H_{\text{p}}})},
\end{equation}
instead of $\rho_{\text{p}}(\infty)\propto e^{-\beta (H_{\text{p}}+H_{\text{c}})}$. This result is quite different from that of the qubit-based temperature sensing case~\cite{PhysRevResearch.3.043039}, in which the effect of frequency renormalization is fully included in the steady state of the probe. From the above Eq.~(\ref{eq:eq12}), one can easily find $\pmb{d}_{\text{M}}(\infty)=(0,0)^{\text{T}}$ and
\begin{equation}\label{eq:eq13}
\pmb{\sigma}_{\text{M}}(\infty)=\coth\Big{(}\frac{\omega_{0}}{2T}\Big{)}\left(
\begin{array}{cc}
\omega_{0}^{-1} & 0 \\
0 & \omega_{0} \\
\end{array}
\right)
\end{equation}
are independent of the details of the spectral density, which leads to $\mathcal{F}_{\lambda}^{\text{M}}(\infty)=0$. It is necessary to point out that $\pmb{d}_{\text{M}}(\infty)=(0,0)^{\text{T}}$ and Eq.~(\ref{eq:eq13}) can be reproduced by directly calculating the equilibration dynamics of the probe under the Markovian approximation (see Appendix B for details).

However, such a canonical thermalization totally breaks down, namely, $\rho_{\text{p}}(\infty)\neq\rho_{\text{p}}^{\text{G}}$, in the strongly non-Markovian regime where the noncanonical equilibrium state appears. As demonstrated in Refs.~\cite{PhysRevA.90.032114,doi:10.1063/1.4722336}, the emergence of a noncanonical distribution is commonly linked to the existence of probe-reservoir correlations, which implies there exists information exchange between the probe and the thermal reservoir. This result suggests, in the non-Markovian case, the long-time steady state of the probe will rely on not only the reservoir's temperature, but also the details of the spectral density, resulting in $\mathcal{F}_{\lambda}(\infty)>0$.

To check the above analysis, in Figs.~\ref{fig:fig5}(a) and ~\ref{fig:fig5}(b), we display $\sigma_{xx}(\infty)$ and $\sigma_{pp}(\infty)$ versus the temperature of the quantum reservoir from using the non-Markovian and the Markovian methods. One can find that the non-Markovian results depart from the results predicted by the canonical Gibbs state when temperature is very low, for example, $T/\omega_{0}\in(0,2]$ in Fig.~\ref{fig:fig5}(b). This result means the breakdown of the Markovian approximation and the appearance of the noncanonical equilibration in the low-temperature regime. In Figs.~\ref{fig:fig5}(c) and ~\ref{fig:fig5}(d), $\mathcal{F}_{\lambda}(\infty)$ is plotted as a function of the temperature. One can see $\mathcal{F}_{\lambda}(\infty)>0$ if the temperature is very low. However, with increasing temperature, the non-Markovianity becomes ignorable and $\mathcal{F}_{\lambda}(\infty)$ gradually vanishes. In this sense, the non-zero residual QFI at low temperature stems from non-Markovian effects which cannot be predicted by the previous Markovian approaches~\cite{PhysRevA.97.012125,PhysRevA.102.012223,PhysRevLett.98.160401,PhysRevA.95.012109}. This result means the non-Markovianity is not only a mathematical concept, but also a valuable resource to improve the sensing precision.

\subsection{Enhanced sensing performance by adding an external driving field}\label{sec:sec3d}

\begin{figure}
\includegraphics[angle=0,width=0.485\textwidth]{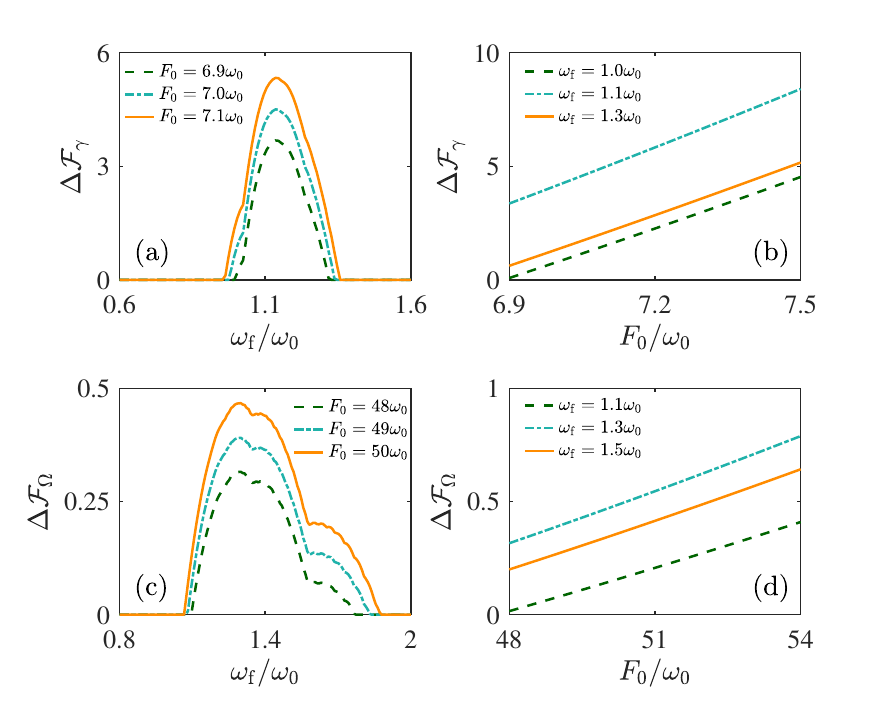}
\caption{\label{fig:fig6} $\Delta\mathcal{F}_{\gamma}$ is plotted as a function of (a) the driving frequency $\omega_{\rm f}$ and (b) the driving amplitude $F_{0}$. $\Delta\mathcal{F}_{\Omega}$ vs (c) the driving frequency $\omega_{\rm f}$ and (d) the driving amplitude $F_{0}$. Parameters are chosen as $\omega_{0}=0.5~\text{THz}$, $\Omega=10\omega_0$, $\gamma=3\omega_0$, $r=5\omega_0$, $\alpha=0$, and $\theta=\theta_{\text{opt}}$.}
\end{figure}

In this subsection, we present a dynamical steer protocol to improve the sensing performance by using an external driving field, which is solely applied to the probe. To this aim, in the original Hamiltonian of the probe, we add the following time-dependent term~\cite{GOLOVINSKI2020126203,doi:10.1063/1.3078024}
\begin{equation}
H_{\text{f}}(t)=-F_{0}\sin(\omega_{\text{f}}t)x,
\end{equation}
where $F_{0}$ and $\omega_{\text{f}}$ are the driving amplitude and frequency, respectively. Using the method of deriving the first two moments given in Appendix A, the QFI under driving can be obtained without difficulties. To quantify the influences of the continuous-wave driving field on the sensing precision, we define the following quantity:
\begin{equation}
\Delta\mathcal{F}_{\lambda}\equiv\max_{t}\mathcal{F}_{\lambda}(F_{0},\omega_{\text{f}})-\max_{t}\mathcal{F}_{\lambda}(F_{0}=0,\omega_{\text{f}}=0).
\end{equation}
As long as $\Delta\mathcal{F}_{\lambda}>0$, one can conclude that the external driving field plays a positive role in our sensing performance. As shown in Fig.~\ref{fig:fig6}, via applying an external driving field, the sensing precision can be effectively improved. Moreover, by adjusting either the driving amplitude $F_{0}$ or the frequency $\omega_{\text{f}}$, the effect of the external driving field can be further optimized. From Fig.~\ref{fig:fig6}, one can see that the constructive effect generated by optimizing the driving amplitude is rather robust: $\Delta\mathcal{F}_{\lambda}$ scales linearly with the increase of the value of $F_{0}$. However, when the driving frequency $\omega_{\text{f}}$ is neither too high nor too low, the dynamical steer effect induced by varying the driving frequency $\omega_{\text{f}}$ becomes negligible.

\section{Conclusion}\label{sec:sec4}

In summary, we employ a harmonic oscillator, which is initially prepared in a squeezed state, acted as a probe to estimate the parameters of the spectral density of a bosonic reservoir. Going beyond the usual Markovian treatment, a SQL-type scaling relation, which can be further optimized by increasing the proportion of squeezing in the initial quantum resource, is revealed. To maximize the sensing precision, we analyze the influences of the form of probe-reservoir interaction, the non-Markovianity as well as an external driving field on the sensing performance. It is found that the sensing sensitivity the can be significantly improved by including $px_{k}$-type coupling, which is commonly neglected in the standard Caldeira-Leggett model. At low temperature, we find that the non-Markovianity can lead to a noncanonical equilibrium state, which contains information of the spectral density under the influence of decoherence. Using such a non-Markovian effect, our sensing scheme still works in the long-encoding-time regime where the Markovian one completely breaks down. Moreover, we propose a dynamical steer protocol, in which an external driving field is applied to the probe, to boost the sensing outcome. Our results presented in this paper may provide some theory supports for designing a high-precision quantum sensor. Furthermore, due to the importance of the spectral density in the theory of open quantum systems, we expect our results to be of interest for understanding and controlling the decoherence.

\section{Acknowledgments}

The authors thank Dr. Si-Yuan Bai, Dr. Chong Chen, Professor Jun-Hong An and Professor Hong-Gang Luo for many fruitful discussions. The work was supported by the National Natural Science Foundation (Grants No. 11704025 and No. 12047501).

\begin{widetext}

\section{Appendix A: The exact expressions of $\pmb{d}$ and $\pmb{\sigma}$}

In this appendix, we would like to show how to derive exact expressions of $\pmb{d}$ and $\pmb{\sigma}$. Using the Heisenberg equation of motion of $\mathcal{\dot{O}}=-i[\mathcal{O},H]$, one can find that the equations of motion of $x$, $x_{k}$, $p$, and $p_{k}$ are given by ($m=1$)
\begin{equation}\label{eq:eq16}
\dot{x}=p+\sin\theta\bigg{(}\sum_{k}c_{k}x_{k}+\sum_{k}\frac{c_{k}^{2}}{m_{k}\omega_{k}^{2}}\mathcal{S}\bigg{)},
\end{equation}
\begin{equation}\label{eq:eq17}
\dot{p}=-\omega_{0}^{2}x-\cos\theta\bigg{(}\sum_{k}c_{k}x_{k}+\sum_{k}\frac{c_{k}^{2}}{m_{k}\omega_{k}^{2}}\mathcal{S}\bigg{)},
\end{equation}
\begin{equation}\label{eq:eq18}
\dot{x}_{k}=\frac{p_{k}}{m_{k}},~~~\dot{p}_{k}=-m_{k}\omega_{k}^{2}x_{k}-c_{k}\mathcal{S}.
\end{equation}
From Eq.~(\ref{eq:eq18}), one can find that the formal solution of $x_{k}$ is
\begin{equation}
x_{k}(t)=x_{k}(0)\cos(\omega_{k}t)+\frac{p_{k}(0)}{m_{k}\omega_{k}}\sin(\omega_{k}t)-\sum_{k}\frac{c_{k}}{m_{k}\omega_{k}}\int_{0}^{t}d\tau\sin[\omega_{k}(t-\tau)]\mathcal{S}(\tau),
\end{equation}
where $\mathcal{S}(t)=x(t)\cos\theta+p(t)\sin\theta$. Substituting the above formal solution of $x_{k}$ into Eqs.~(\ref{eq:eq16}) and ~(\ref{eq:eq17}), one can find the following integro-differential equation for the position operator :
\begin{equation}\label{eq:eq20}
\ddot{x}(t)+\omega_{0}^{2}x(t)+\bigg{(}\cos\theta\frac{d}{dt}-\sin\theta\frac{d^{2}}{dt^{2}}\bigg{)}\int_{0}^{t}d\tau \mathcal{Z}(t-\tau)\mathcal{S}(\tau)=\sin\theta \mathcal{\dot{R}}(t)-\cos\theta\mathcal{R}(t),
\end{equation}
where
\begin{equation}
\mathcal{Z}(t)\equiv\sum_{k}\frac{c_{k}^{2}}{m_{k}\omega_{k}^{2}}\cos(\omega_{k}t),~~~\mathcal{R}(t)\equiv\sum_{k}c_{k}\bigg{[}x_{k}(0)\cos(\omega_{k}t)+\frac{p_{k}(0)}{m_{k}\omega_{k}}\sin(\omega_{k}t)\bigg{]},
\end{equation}
are, respectively, the so-called damping kernel and environment-induced stochastic force. In this paper, we consider the reservoir to be initially prepared as $\rho_{\text{r}}^{\text{G}}=e^{-\beta H_{\text{r}}}/\text{Tr}(e^{-\beta H_{\text{r}}})$. This assumption leads to $\langle\mathcal{R}(t)\rangle_{\text{r}}=0$ and the symmetrized environmental correlation function is given by
\begin{equation}
\mathcal{C}(t)\equiv\frac{1}{2}\Big{[}\langle\mathcal{R}(t)\mathcal{R}(0)\rangle_{\text{r}}+\langle\mathcal{R}(0)\mathcal{R}(t)\rangle_{\text{r}}\Big{]}=\sum_{k}\frac{c_{k}^{2}}{2m_{k}\omega_{k}^{2}}\coth\Big{(}\frac{\beta\omega_{k}}{2}\Big{)}\cos(\omega_{k}t),
\end{equation}
where $\langle\mathcal{O}\rangle_{\text{r}}\equiv \text{Tr}_{\text{b}}(\rho_{\text{r}}^{\text{G}}\mathcal{O})$. For the Ohmic spectral density considered in this paper, one can find $\mathcal{Z}(t)=\gamma\Omega e^{-\Omega t}$ and
\begin{equation}
\begin{split}
\mathcal{C}(t)=&\frac{2\gamma}{\beta}e^{-\Omega t}+\frac{4\gamma\Omega^{2}}{\beta}\sum_{n=1}^{\infty}\frac{\nu_{n}e^{-\nu_{n}t}-\Omega e^{-\Omega t}}{\nu_{n}^{2}-\Omega^{2}}=\frac{2\gamma}{\beta}e^{-\Omega t}-\frac{4\gamma\Omega^{2}}{\beta}\sum_{n=1}^{\infty}\frac{\Omega e^{-\Omega t}}{\nu_{n}^{2}-\Omega^{2}}+\frac{4\gamma\Omega^{2}}{\beta}\sum_{n=1}^{\infty}\frac{\nu_{n}e^{-\nu_{n}t}}{\nu_{n}^{2}-\Omega^{2}}\\
=&\frac{\gamma}{\beta}\bigg{[}2(1-\Omega)+\beta\Omega^{2}\cot\Big{(}\frac{\beta\Omega}{2}\Big{)}\bigg{]}e^{-\Omega t}+\frac{\gamma\Omega^{2}}{\pi}\bigg{[}\mathbb{H}\Big{(}e^{-\frac{2\pi t}{\beta}},1,1-\frac{\beta\Omega}{2\pi}\Big{)}+\mathbb{H}\Big{(}e^{-\frac{2\pi t}{\beta}},1,1+\frac{\beta\Omega}{2\pi}\Big{)}\bigg{]}e^{-\frac{2\pi t}{\beta}},
\end{split}
\end{equation}
where $\nu_{n}\equiv 2n\pi/\beta$ are the Matsubara frequencies, and $\mathbb{H}(a,b,c)$ denotes the Lerch transcendent function.

Solving the Eq.~(\ref{eq:eq20}) by using Laplace transformation, one can find
\begin{equation}\label{eq:eq24}
x(t)=\mathcal{G}_{1}(t)x(0)+\mathcal{G}_{2}(t)p(0)+\int_{0}^{t}d\tau \mathcal{G}_{3}(t-\tau)\mathcal{R}(\tau),
\end{equation}
\begin{equation}\label{eq:eq25}
p(t)=\mathcal{G}_{4}(t)x(0)+\mathcal{G}_{5}(t)p(0)+\int_{0}^{t}d\tau \mathcal{G}_{6}(t-\tau)\mathcal{R}(\tau).
\end{equation}
Here, $\mathcal{G}_{\alpha}(t)$ with $\alpha=1,2,3,4,5,6$ are determined by the following inverse Laplace transformation
\begin{equation}
\mathcal{G}_{1}(t)=\mathcal{L}^{-1}\Bigg{[}\frac{z+z\cos\theta\sin\theta \tilde{\mathcal{Z}}(z)}{\zeta(z)}\Bigg{]},~~\mathcal{G}_{2}(t)=\mathcal{L}^{-1}\Bigg{[}\frac{1+z\sin^{2}\theta \tilde{\mathcal{Z}}(z)}{\zeta(z)}\Bigg{]},~~\mathcal{G}_{3}(t)=\mathcal{L}^{-1}\Bigg{[}\frac{z\sin\theta-\cos\theta}{\zeta(z)}\Bigg{]},
\end{equation}
\begin{equation}
\mathcal{G}_{4}(t)=\mathcal{L}^{-1}\Bigg{[}\frac{-\omega_{0}^{2}-z\cos^{2}\theta\tilde{\mathcal{Z}}(z)}{\zeta(z)}\Bigg{]},~~\mathcal{G}_{5}(t)=\mathcal{L}^{-1}\Bigg{[}\frac{z-z\sin\theta\cos\theta \tilde{\mathcal{Z}}(z)}{\zeta(z)}\Bigg{]},~~\mathcal{G}_{6}(t)=\mathcal{L}^{-1}\Bigg{[}\frac{-\omega_{0}^{2}\sin\theta-z\cos\theta}{\zeta(z)}\Bigg{]},
\end{equation}
where $\zeta(z)\equiv\omega_0^2+z^2+z\tilde{\mathcal{Z}}(z)(\cos^2\theta+\omega_0^2\sin^2\theta)$ and $\mathcal{Z}(z)=\mathcal{L}[\mathcal{Z}(t)]=\gamma\Omega/(z+\Omega)$. By making use of residue theorem, the inverse Laplace transformation can be exactly worked out:
\begin{equation}
\mathcal{G}_{\alpha}(t)=\sum_{\text{i}\neq\text{j}\neq\text{k}=1}^{3}\frac{G_{\alpha}(z_{\text{i}})e^{z_{\text{i}}t}}{(z_{\text{i}}-z_{\text{j}})(z_{\text{i}}-z_{\text{k}})},
\end{equation}
where $z_{\text{i}}$ are the roots of the cubic polynomial $z^{3}+\Omega z^{2}+[\omega_{0}^{2}+\gamma\Omega(\cos^{2}\theta+\omega_{0}^{2}\sin^{2}\theta)]z+\omega_{0}^2\Omega=0$, and $G_{\alpha}(z)$ are given by
\begin{equation}
G_{1}(z)=z(z+\Omega)+z\gamma\Omega\cos\theta\sin\theta,~~G_{2}(z)=z+\Omega+z\gamma\Omega\sin^{2}\theta,~~G_{3}(z)=(z+\Omega)(z\sin\theta-\cos\theta),
\end{equation}
\begin{equation}
G_{4}(z)=-\omega_{0}^{2}(z+\Omega)-z\gamma\Omega\cos^{2}\theta,~~G_{5}(z)=z(z+\Omega)-z\gamma\Omega\cos\theta\sin\theta,~~G_{6}(z)=-(z+\Omega)(\omega_{0}\sin\theta+z\cos\theta).
\end{equation}

Using Eqs.~(\ref{eq:eq24}), ~(\ref{eq:eq25}), and some straightforward calculations, we can finally obtain the exact expressions of the first two momentums as $d_{x,p}=\mathcal{G}_{1,4}(t)\langle x(0)\rangle_{\text{p}}+\mathcal{G}_{2,5}(t)\langle p(0)\rangle_{\text{p}}$ and
\begin{equation}
\sigma_{xx}(t)=\mathcal{G}_1^2(t)\sigma_{xx}(0)+\mathcal{G}_2^2(t)\sigma_{pp}(0)+2\mathcal{G}_1(t)\mathcal{G}_2(t)\sigma_{xp}(0)+2\int_{0}^{t}d\tau\int_{0}^{t}d\tau'\mathcal{G}_3(t-\tau)\mathcal{G}_3(t-\tau')\mathcal{C}(\tau-\tau'),
\end{equation}
\begin{equation}
\sigma_{pp}(t)=\mathcal{G}_4^2(t)\sigma_{xx}(0)+\mathcal{G}_5^2(t)\sigma_{pp}(0)+2\mathcal{G}_4(t)\mathcal{G}_5(t)\sigma_{xp}(0)+2\int_{0}^{t}d\tau\int_{0}^{t}d\tau'\mathcal{G}_6(t-\tau)\mathcal{G}_6(t-\tau')\mathcal{C}(\tau-\tau'),
\end{equation}
\begin{equation}
\begin{split}
\sigma_{xp}(t)=&\mathcal{G}_1(t)\mathcal{G}_4(t)\sigma_{xx}(0)+\mathcal{G}_2(t)\mathcal{G}_5(t)\sigma_{pp}(0)+2[\mathcal{G}_2(t)\mathcal{G}_4(t)+\mathcal{G}_1(t)\mathcal{G}_5(t)]\sigma_{xp}(0)\\
&+2\int_{0}^{t}d\tau\int_{0}^{t}d\tau'\mathcal{G}_3(t-\tau)\mathcal{G}_6(t-\tau')\mathcal{C}(\tau-\tau'),
\end{split}
\end{equation}
where $\langle\mathcal{O}\rangle_{\text{p}}\equiv\text{Tr}_{\text{p}}[\rho_{\text{p}}(0)\mathcal{O}]$ and $\sigma_{ij}(0)$ denotes the initial covariant matrix with respect to $\rho_{\text{p}}(0)$.

\section{Appendix B: Markovian results}

\begin{figure*}
\includegraphics[angle=0,width=0.975\textwidth]{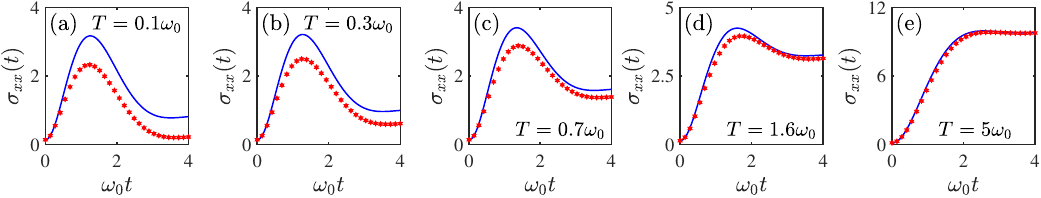}
\caption{\label{fig:fig7} The dynamics of the second moment $\sigma_{xx}(t)$ with different temperatures: (a) $T=0.1\omega_0$, (b) $T=0.3\omega_0$, (c) $T=0.7\omega_0$, (d) $T=1.6\omega_0$, (e) $T=5\omega_0$. The red hexagons are the Markovian results, while the blue solid lines are obtained by the non-Markovian method. Other parameters are chosen as $\omega_{0}=1~\text{THz}$, $\Omega=10\omega_0$, $\gamma=\omega_0$, $r=\omega_0$, $\alpha=0$, and $\theta=0$.}
\end{figure*}

In the case of $\gamma,\omega_0\ll\text{min}\{\Omega,2\pi T\}$, Refs.~\cite{Breuer,PhysRevA.102.022228} provided a Markovian approximate expressions of $\mathcal{G}_{\alpha}(t)$ in the case of $\theta=0$ as follows:
\begin{equation}\label{eq:eq34}
\mathcal{G}_{1}^{\text{M}}(t)\simeq\bigg{[}\frac{\kappa}{\Lambda}\sin(\Lambda t)+\cos(\Lambda t)\bigg{]}e^{-\kappa t},~~\mathcal{G}_{2}^{\text{M}}(t)\simeq\frac{1}{\Lambda}\sin(\Lambda t)e^{-\kappa t},~~
\mathcal{G}_{3}^{\text{M}}(t)\simeq-\frac{1}{\Lambda}\sin(\Lambda t)e^{-\kappa t},
\end{equation}
\begin{equation}\label{eq:eq35}
\mathcal{G}_{4}^{\text{M}}(t)\simeq-\frac{\omega_0^2}{\Lambda}\sin(\Lambda t)e^{-\kappa t},~~
\mathcal{G}_{5}^{\text{M}}(t)\simeq\cos(\Lambda t)e^{-\kappa t}-\frac{\kappa}{\Lambda}\sin(\Lambda t)e^{-\kappa t},~~
\mathcal{G}_{6}^{\text{M}}(t)\simeq-\cos(\Lambda t)e^{-\kappa t}+\frac{\kappa}{\Lambda}\sin(\Lambda t)e^{-\kappa t},
\end{equation}
where $\kappa\equiv\gamma/2$ and $\Lambda^{2}\equiv\omega_{0}^{2}-\kappa^{2}$. On the other hand, in the Markovian treatment, the environmental correlation function $\mathcal{C}(\tau-\tau')$ reduces to a Dirac-$\delta$ function, i.e., $\mathcal{C}_{\text{M}}(\tau-\tau')\simeq4\gamma T\delta(\tau-\tau')$. With these above approximate expressions of $\mathcal{G}_{\alpha}^{\text{M}}(t)$ and $\mathcal{C}_{\text{M}}(\tau-\tau')$ at hand, the first two momentums under the Markovian approximation can easily be obtained. In Fig.~\ref{fig:fig7}, we display $\sigma_{xx}(t)$ from both the Markovian and non-Markovian methods. Good agreement is found between results from the above two different approaches at high temperature, say $T=5\omega_{0}$ in Fig.~\ref{fig:fig7}(e). However, as the environmental temperature decreases, the non-Markovian effect becomes strong. At low temperature, e.g., $T=0.1\omega_{0}$ in Fig.~\ref{fig:fig7}(a), a relatively large deviation is found, which means the breakdown of the Markovian approximation. Moreover, using Eqs.~(\ref{eq:eq34}), ~(\ref{eq:eq35}), and the Dirac-$\delta$-type environmental correlation function, one can easily derive the expressions of the first two momentums in the long-encoding-limit, which recovers $\pmb{d}_{\text{M}}(\infty)=(0,0)^{\text{T}}$ and Eq.~(\ref{eq:eq13}) in the main text. This result demonstrate that the probe experiences a canonical thermalization under the Markovian approximation, which is consistent with Refs.~\cite{PhysRevE.90.022122,PhysRevA.89.012128,PhysRevA.90.032114,doi:10.1063/1.4722336}.

\section{Appendix C: The derivation of Eq.~(\ref{eq:eq8})}

In this appendix, we show the details of deriving the Eq.~(\ref{eq:eq8}). By making the Lagrange interpolation method, a smooth function $f_{\lambda}=f(\lambda)$, which is defined in a tiny interval $\lambda\in[\lambda_{\text{min}},\lambda_{\text{max}}]$, can be approximately expressed as a sum of polynomials,
\begin{equation}
f(\lambda)\simeq\sum_{\text{n}=0}^{N}f(\lambda_{\text{n}})\mathbb{L}_{\text{n}}(x),
\end{equation}
where $\lambda_{\text{min}}=\lambda_{0}<\lambda_{1}<...<\lambda_{N}=\lambda_{\text{max}}$ with $\lambda_{\text{n}+1}=\lambda_{\text{n}}+\delta$;  $\delta=(\lambda_{\text{max}}-\lambda_{\text{min}})/N$ are $(N+1)$ uniformly spaced nodes; and
\begin{equation}
\mathbb{L}_{\text{n}}(x)\equiv\prod_{\text{m}\neq \text{n}}\frac{x-x_{\text{m}}}{x_{\text{n}}-x_{\text{m}}}
\end{equation}
is the so-called Lagrange multiplier function. Taking $N=4$ as an example, we have
\begin{equation}
\begin{split}
f(\lambda)\simeq&\frac{(\lambda-\lambda_{1})(\lambda-\lambda_{2})(\lambda-\lambda_{3})(\lambda-\lambda_{4})}{(\lambda_{0}-\lambda_{1})(\lambda_{0}-\lambda_{2})(\lambda_{0}-\lambda_{3})(\lambda_{0}-\lambda_{4})}f(\lambda_{0})+\frac{(\lambda-\lambda_{0})(\lambda-\lambda_{2})(\lambda-\lambda_{3})(\lambda-\lambda_{4})}{(\lambda_{1}-\lambda_{0})(\lambda_{1}-\lambda_{2})(\lambda_{1}-\lambda_{3})(\lambda_{1}-\lambda_{4})}f(\lambda_{1})\\
&+\frac{(\lambda-\lambda_{0})(\lambda-\lambda_{1})(\lambda-\lambda_{3})(\lambda-\lambda_{4})}{(\lambda_{2}-\lambda_{0})(\lambda_{2}-\lambda_{1})(\lambda_{2}-\lambda_{3})(\lambda_{2}-\lambda_{4})}f(\lambda_{2})+\frac{(\lambda-\lambda_{0})(\lambda-\lambda_{1})(\lambda-\lambda_{2})(\lambda-\lambda_{4})}{(\lambda_{3}-\lambda_{0})(\lambda_{3}-\lambda_{1})(\lambda_{3}-\lambda_{2})(\lambda_{3}-\lambda_{4})}f(\lambda_{3})\\
&+\frac{(\lambda-\lambda_{0})(\lambda-\lambda_{1})(\lambda-\lambda_{2})(\lambda-\lambda_{3})}{(\lambda_{4}-\lambda_{0})(\lambda_{4}-\lambda_{1})(\lambda_{4}-\lambda_{2})(\lambda_{4}-\lambda_{3})}f(\lambda_{4}).
\end{split}
\end{equation}
Assuming $f(\lambda)$ is differentiable in the interval of $[\lambda_{2}-2\delta,\lambda_{2}+2\delta]$, then the first-order derivative evaluated at $\lambda=\lambda_{2}$ can be approximately written as
\begin{equation}
\begin{split}
f'(\lambda_{2})\simeq\frac{d}{d\lambda}f(\lambda)\bigg{|}_{\lambda=\lambda_{2}}=\frac{1}{12\delta}[f(\lambda_{0})-8f(\lambda_{1})+8f(\lambda_{3})-f(\lambda_{4})],
\end{split}
\end{equation}
which recovers Eq.~(\ref{eq:eq8}) in the main text.
\end{widetext}

\bibliography{reference}

\end{document}